\documentclass[nofootinbib,preprint,aps]{revtex4} 
\usepackage{graphicx}
\begin{document}
\title{Small Lorentz violations in quantum gravity: do they lead to
  unacceptably large effects?}

\author{Rodolfo Gambini, Saeed Rastgoo}
\address{Instituto de F\'{\i}sica, Facultad de Ciencias, 
Universidad
de la Rep\'ublica, Igu\'a 4225, CP 11400 Montevideo, Uruguay}

\author{Jorge Pullin}
\address{Department of Physics and Astronomy, 
Louisiana State University, Baton Rouge,
LA 70803-4001}

\date{February 21 2011}

\begin{abstract}
  We discuss the applicability of the argument of Collins, P\'erez,
  Sudarsky, Urrutia and Vucetich to loop quantum gravity. This
  argument suggests that Lorentz violations, even ones that only
  manifest themselves at energies close to the Planck scale, have
  significant observational consequences at low energies when one
  considers perturbative quantum field theory and renormalization. We
  show that non-perturbative treatments like those of loop quantum
  gravity may generate deviations of Lorentz invariance of a different
  type than those considered by Collins et al. that do not necessarily
  imply observational consequences at low energy.
\end{abstract}

\maketitle
Over time, the notion that quantum gravity may impose some level of
discreteness to space-time at the Planck scale, has been put forward
in many contexts (see \cite{loll} for a reviews). In loop quantum 
gravity (LQG) and other approaches it is many times ---though not
universally--- asserted that there is a fundamental level of
discreteness. For instance, discretized models of spherically
symmetric space-times seem to impose a Planck scale lattice at a
fundamental level \cite{spherical}. So do models of parameterized
field theory \cite{laddha}.  In spin foam approaches it is sometimes
assumed that one will not take the limit in which the spacing in the
foam goes to zero \cite{oriti} but rather look for coarse graining of
the physical quantities at large distances compared to the Planck
scale. The presence of a fundamental discreteness does not necessarily
imply that there exist violations of the local Lorentz invariance. For
instance, many spin foam models implement local Lorentz invariance
exactly. However, some of the models of discreteness seem to violate
local Lorentz invariance.  Lorentz violating theories have also
received recent attention in relation to Ho\v{r}ava's proposal of
gravity at the Lifshitz point \cite{horava}

In such contexts an argument due to Collins, P\'erez, Sudarsky,
Urrutia and Vucetich\cite{collins,collins2} seems to be quite
relevant. These authors have studied what happens to perturbative
quantum field theories that violate Lorentz invariance. They notice
that when one renormalizes such theories even minute Lorentz
violations that manifest themselves only at the Planck scale become
quite relevant. When one regularizes and renormalizes there are
certain terms that would be divergent that just happen to be zero due
to Lorentz invariance. When the latter is broken, even by amounts that
are very small at low energies, the regularization and renormalization
process can render results that violate Lorentz invariance in amounts
not suppressed by the energy.  These effects are large 
enough to render the resulting theories
experimentally not viable.

Should such an argument imply that the Lorentz violations that may
arise in LQG due to its discreteness render the
theory not viable experimentally?  There is a bit of a gap in the
argument. The original argument talks about building perturbative
quantum field theories that start being Lorentz non-invariant and
relies on the use of renormalization.  But LQG is
non-perturbative and in most calculations one does not need to
renormalize since the theory is finite. Does this imply that the
argument does not apply in this case? Not necessarily. LQG
aspires to make contact for low energies with ordinary
perturbative quantum field theory. How is such a contact going to
accommodate potential Lorentz violations without running into the
problems pointed out by Collins et al.?

To illustrate the problem, we will consider a simplified situation,
discussed by Collins et al. \cite{collins2}, the case of a Yukawa
interaction of a scalar field and a Fermion. However, we will not
introduce by hand Lorentz violating terms that go as $E/E_{\rm
  Planck}$. We will put the theory on a lattice and we will assume the
lattice spacing is small compared to particle physics lengths but
large or comparable to the Planck scale, and that will be our source
of Lorentz non-invariance. This is the situation one faces, for
instance, in spherical models of LQG, which leads to
similar types of propagators as those of the model we consider. We
will not take the limit in which the lattice spacing goes to zero (if
we did we would reproduce the results of Collins et al. without
Lorentz violating terms, just with a different regularization, as they
used Pauli--Villars to regularize).

On the lattice the propagators differ from the continuum ones. The one
for the Fermion is,
\begin{equation}
  S(k) = \frac{m-i\sum_{j=1}^3 \gamma^j a^{-1} \sin\left(a
      k_j\right)-i\gamma^0 \left(b a\right)^{-1}\sin\left(b\, a
      k_0\right)}
{m^2 +a^{-2} \sum_{j=1}^3 \left(2 -2\cos^2\left(a k_j\right)\right)+\left(b\,
    a\right)^{-2} \left(2-2\cos^2\left(b\, a k_0\right)\right)}.
\end{equation}
We are considering the Euclidean case as is common in lattice field
theory and we are allowing different lattice spacings in space $a$ and
time, $b\,a$ with $b$ an arbitrary factor that tends to one in
the continuum limit; unequal spacings in space and time might occur
in LQG in the canonical approach). For
the scalar we have ,
\begin{equation}
  G(k,m) = \frac{1}{m^2 +a^{-2} 
\sum_{j=1}^3 \left(2 -2\cos^2\left(a k_j\right)\right)+\left(b\,a\right)^{-2}
\left(2 -2\cos^2\left(b\,a k_0\right)\right)},
\end{equation}
The derivation of these propagators can be seen in reference
\cite{smit}. In the limit $a\to 0$ one recovers the usual expressions
for the propagator.  If one computes the correlation length between
two lattice sites $\vec{n}_1$ and $\vec{n}_2$,
$\gamma(\vec{n}_1,\vec{n_2})$, given by taking the Fourier transform
of $G(k,m)$ it will in general go as $G\sim
\exp(-t/\gamma(\vec{n}_1,\vec{n}_2))$ with $t$ the distance between
the two lattice site will in general depend on the choice of lattice
sites.  However, in the continuum limit it becomes isotropic, taking
the quotient of two $\gamma$'s in arbitrary directions $
\frac{\gamma(\vec{n}_1,\vec{n}_2)} {\gamma(\vec{n}_1,\vec{n}_3)} \sim
1+O(m^2 a^2)$.

This indicates that the lattice treatment restores the invariances of
the continuum for finite values of $a$ in a continuous way. It gives a
hint that in general one will not have large departures from the 
continuum for small lattice spacings, as we will see.

\begin{figure}[h]
\includegraphics[height=3cm]{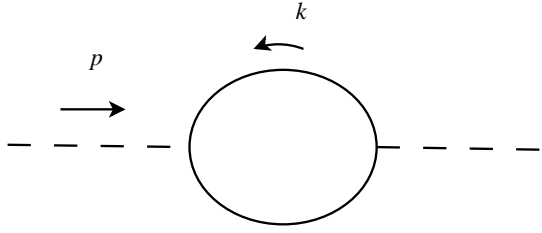}
\caption{Feynman diagram for the lowest order self-energy. The diagram
  includes an unbound integral in $k$ that can, for energy dependent
  Lorentz violations, lead to large observable effects even if the
  Lorentz violations are small for low energies.}
\label{fig1}
\end{figure}


The sum over irreducible one particle two-point graphs, called the
self-energy $\Pi(p)$, yields the corrections to the propagation of the
scalar field. Following Collins et al., we study its value for momenta
and mass much smaller than a cutoff $\Lambda$ (that in the lattice
case will be related to the inverse of $a$) and to one loop order, as
shown in figure 1, which we denote by $\Pi_1(p)$. With no cutoff the
diagram is quadratically divergent.  One has that,
\begin{equation}
\Pi_1(p)= A+p^2 B+p^\mu p^\nu W_\mu W_\nu \tilde{\xi}+\Pi^{(LI)}(p^2) +
O(p^4/\Lambda^2).
\end{equation}
with $W_\mu$ a timelike four vector.  In the continuum the first two
terms diverge and the divergence is absorbed in a redefinition of the
mass and wavefunction. The fourth term is Lorentz invariant and
finite. The fifth term is Lorentz violating but suppressed. The third
term may contain unsuppressed Lorentz violations. To obtain the third
term we take the self-energy and differentiate it twice, $
\xi=\frac{\partial^2 \Pi_1(p)}{\partial \left(p^0\right)^2}
-\frac{\partial^2 \Pi_1(p)}{\partial \left(p^1\right)^2},$
and evaluate it at $p=0$. Such a quantity obviously vanishes for a
Lorentz invariant theory (differences in signs with
Collins et al. are due to us working in the Euclidean case). 
Again following Collins et al. in the continuum one has,
\begin{equation}
  \tilde{\xi}= -\frac{ig^2}{\pi^4} \int d^4k
  \frac{\left[-\left(k^0\right)^2+\left(k^1\right)^2\right]
\left(k^2-3m^2\right)}{\left(k^2+m^2\right)^4}
\end{equation}
and in the lattice the analogous expression is given by,
\begin{eqnarray}
  \tilde{\xi}_a&=&
  \int_{-\pi/\left(b\,a\right)}^{\pi/\left(b\,a\right)} 
dk_0\int_{-\pi/a}^{\pi/a}
dk_1 \int_{-\pi/a}^{\pi/a} dk_2\int_{-\pi/a}^{\pi/a} dk_3
\left(\frac{\sum_j\sin^2\left(a k_j\right)}{a^2}
+\frac{\sin^2\left(b\,a k_0\right)}{b^2a^2}+m^2\right)^{-4}\nonumber\\
&&\times\left[
16\left(\frac{\sum_j\sin^2\left(a k_j\right)}{a^2}+
\frac{\sin^2\left(b\,a k_0\right)}{b^2a^2}
+m^2\right)
\left(\frac{\sin^2\left(2 b\,a k_0\right)}{4b^2a^2}
-\frac{\sin^2\left(2 a k_1\right)}{4a^2}\right)\right.\nonumber\\
&&+4\left(\sin^2\left(b\,a k_0\right)-\sin^2\left(a k_1\right)\right)
\left(\frac{\sum_j\sin^2\left(a k_j\right)}{a^2}
+\frac{\sin^2\left(a k_0\right)}{b^2a^2}+m^2\right)^{2}\nonumber\\
&&+32\left(\frac{\sin^2\left(2 b\,a k_0\right)}{4b^2a^2}
-\frac{\sin^2\left(2 a k_1\right)}{4a^2}\right)
\left(m^2
-\frac{\sum_j\sin^2\left( a k_j\right)}{a^2}
-\frac{\sin^2\left(b\, a k_0\right)}{b^2a^2}\right)\nonumber\\
&&\left.-8 \left(\cos\left(2 b\,a k_0\right)-\cos\left(2 a k_1\right)\right) 
\left(m^4-\left(\frac{\sum_j\sin^2\left( a k_j\right)}{a^2}+
\frac{\sin^2\left(b\, a
        k_0\right)}{b^2 a^2}\right)^2\right)
\right]
\end{eqnarray}
where one should strictly have a sum over $k$ but it can be
approximated with an integral with a cutoff of order $1/a$, that is,
each spatial integral goes from $-\pi/a$ to $\pi/a$ and similarly
the time integral goes from $-\pi/(b\,a)$ to $\pi/(b\,a)$. 

Collins et al. showed that in the continuum that $\tilde{\xi}$ is
finite in the limit when the cutoff goes to infinity if one has
propagators that violate Lorentz invariance even if the violation
occurs only at very high energies. That means one has Lorentz
violations unsuppressed by the energy. Let us study the situation in
the lattice.  If one has a finite lattice spacing and computes the
integral (7), if one takes the same spacing in space and in time, the
integral vanishes by symmetry reasons. If one takes different spacings
in space and time (but such that the limit is isotropic, for instance
$b=1+\mu a$ with $\mu$ a parameter with dimensions of mass) 
one gets a finite result, that in the limit,
\begin{equation}
  \lim_{a\to 0} \xi_a = 0,
\end{equation}
and if one expands the finite result in powers of $a$, one gets that
the leading contribution is of the order of $a\,\mu$. 
That is, there
is a departure from the Lorentz invariant case, but if the lattice
spacing is small the departure is small unlike the result of Collins
et al. It is important to notice the order in which limits are 
taken. If one takes $a\to 0$ before computing the integral one is back
to the continuum Lorentz-invariant calculation. 

The central message seems to be that the types of approaches advocated
in LQG, where one considers lattice spacings that are
small but non-vanishing, may not lead to the type of Lorentz-violating
contributions that were considered by Collins et al.

The model we are studying has some imperfections as a complete
illustration of the situation one faces in LQG. First
of all, we have not done a LQG treatment of the
model, but only mimicked the situation using a lattice. Second, we
used a space-time lattice. At first this may appear quite problematic:
in LQG one may work in a canonical framework. Should
we not have used a lattice only in space?  The propagators with a
spatial lattice and continuum time would have denominators
$\left(p^0\right)^2+\sum_{i=1}^3\sin\left(a p^i\right)^2/a^2$.  The
integral in the $p^0$ variable is now computed on an unbound domain
and leads again to the types of effects Collins et al. pointed out. Is
this a problem? We do not necessarily think so. LQG
being a canonical approach, it has the ``problem of time". That is,
the parameter associated with the orbits of the Hamiltonian
constraint, which is continuous, is not to be identified as time. One
possible way out of the problem of time is to do a relational
treatment where one picks one of the variables as a quantum clock
\cite{time}.  Such clocks may have continuous spectrum, but due to
their quantum nature they have uncertainties associated with
them. That situation is not modeled well by the discussion of the
current paper, but we would like to argue that the dispersion in the
clock variable may cure the divergences introduced by the $p^0$ terms and
eliminates any large departure from Lorentz invariance. Let us sketch
how this works. In quantum gravity time will be given by an operator
$T(x^0)$. The propagator will be the ordinary one multiplied a couple
of probability distributions,
\begin{equation}
  D\left(T,\vec{x},T',\vec{x}'\right)=\int_{-\infty}^\infty dt\,dt'
  D\left(t,\vec{x},t',\vec{x}'\right) 
{\cal P}\left(t,T\right){\cal P}\left(t',T'\right),
\end{equation}
where the probability ${\cal P}\left(t,T\right)$ is the probability
that the real clock variable takes a value $T$ when the ideal
parameter is $t$. In situations where there is a well defined notion
of time one expects ${\cal P}(t,T)$ to be a peaked function close to a
Dirac delta (the ideal and real time correlate well). The extra
integral of the peaked function cuts off the high frequency
contributions in zeroth component of the momentum in the
propagator. The others are cut off because of the spatial
lattice. Therefore the resulting space-time propagator is an ordinary
(non distributional) function. As an example if one takes ${\cal P}$
to be a step function regularization of the delta of width $\sigma$
centered in $t$ and $t'$, one gets,
\begin{equation}
D\left(T,\vec{x},T',\vec{x}'\right)    =
\int_{-\pi/a}^{\pi/a} d^3p \frac{e^{i\vec{p}\cdot \vec{x}}}
{2\omega_a} \frac{\sin^2\left(\omega_a \sigma\right)}{\omega_a^2
  \sigma^2} e^{-i\omega_a \vert T-T'\vert},
\end{equation}
where $\omega_a=\sqrt{m^2 +\sum_j\frac{\sin^2\left(a
      p_j\right)}{a^2}}$ and it leads to finite expressions for the
Feynman diagrams. In the limit $a\to 0$, $\sigma\to 0$ one obtains the
usual propagator and, Lorentz invariance of the propagator is
recovered in the limit. In usual quantum gravity scenarios $\sigma$ is
proportional to some power of the Planck length and grows with time.
That means that both the spatial and temporal effects would both go to
zero if one were to take the Planck length to zero. This ensures some
uniformity in the limit for what are two very different
effects. However, it should be emphasized that the above calculation
is just a sketch, with many implications yet to be fleshed out, and at
the present level of development one cannot conclusively state that
the limit will behave as in the lattice example we discussed in
detail. It should also be added that in a field theory like general
relativity one not only expects to include physical clocks to measure
time but also physical measuring rods to measure space (a brief
discussion is in \cite{rods}). If one does that one will produce
smearings similar to the one we just discussed in the other components
of the momentum and this will further help avoid large Lorentz
violations. An analogous situation develops in attempts to regulate
theories using Lorentz-violating non-commutative and fuzzy space
\cite{doplicher}.

It could also be criticized that we worked in the Euclidean theory as
is commonly done in lattice treatments, something one does not expect
to do in the case of LQG. The truth is that any
regularization procedure that violates Lorentz invariance but
preserves the invariance at the dominant order in the ultraviolet
regime will behave like the calculation we carried out. We chose the
Euclidean lattice as an example, but this is not central to the main
argument.  For instance, as an example of a regularization of the type
mentioned, one could have used a Lorentz violating Pauli--Villars
regularization. The usual (Lorentz invariant) Pauli--Villars
regularization consists in including an additional particle with mass
$M$,
\begin{equation}
  \frac{1}{-k_0^2+\vec{k}^2+m^2} -\frac{1}{-k_0^2+\vec{k}^2+M^2}
\end{equation}
and in the limit of large $M$ this regularizes the theory because the
propagator instead of going as $1/k^2$ goes as $1/k^4$ . A Lorentz
violating Pauli--Villars regularization of the propagator could be,
\begin{equation}
  \frac{1}{-k_0^2+\vec{k}^2+m^2}
  -\frac{1}{-k_0^2+\vec{k}^2+\frac{m^4}{\vec{k}^2 +M^2}+M^2}.
\end{equation}
This still regularizes the theory and the Lorentz violating terms go to zero
in the limit in which one removes the regulator. One can check that
computing $\xi$ one gets a small contribution of the order $m^4/M^4$.
Although people commonly perform a Wick rotation when working with
Pauli--Villars regularizations as a matter of convenience, it is not
necessary to work that way, one can work in an entirely Lorentzian
fashion. The Pauli--Villars example is problematic in that outside of
the limit the resulting quantum field theory is ill defined. One can
however think of other regularization schemes that yield a well
defined quantum field theory even when the regulator is not removed.
Also, the Pauli--Villars example suggests that having unbound
integrals in $k_0$ does not automatically lead to the problems
presented in Collins et al. Another example of this is given in
the paper of Reyes, Urrutia and Vergara \cite{reurve}. 
But as our own example shows, one cannot take any lattice
regularization, certain conditions of regularity had to be met. 
This is particularly relevant in the context of gravity where 
irregular lattices are commonly used. We have not studied that case in
detail up to now.

In hindsight, the findings of this paper are not surprising. It is
well known that {\em any} regularization procedure produces, before
taking the limit, propagators that depend on a (dimensionful)
parameter and that are finite. Outside of the limit, all
regularizations produce results that are close to the continuum limit
plus terms that are absorbed in redefinitions of the fundamental
parameters. The important point is there exists continuity in the
regularization parameter. There is abundant literature
(e.g. \cite{abundant}) on what happens outside of the limit and it is
well known that new physics arises. Analyticity properties change at
high energy. But, with the hypotheses mentioned above, there are no
large violations of Lorentz invariance. All this is in line with what
one would expect in a context like LQG.

An additional point is that one should be extra careful when
considering expansions of quantities. For instance if one had
considered the propagator in the lattice and expanded in the limit of
small lattice spacing $a$, and only kept the leading terms, instead of
having trigonometric functions one would have powers of the momentum.
Such terms would lead to large contributions to $\xi$ since they do
have the form considered by Collins et al. However it is clear from
the discussion of the present paper that analyzing things in such a
way is incorrect. We had carried out analyses of that type in
unpublished work \cite{pullin}. It would also be the case if one 
attempted to apply the analysis of Collins et al. to the polymer
propagators computed in \cite{Hossain:2010eb}.



Summarizing, assuming that ---as expected--- LQG could lead to a
finite divergence-free theory whose low energy approximation for
matter fields will be of the form of a regularized version of ordinary
quantum field theories, small potential departures from Lorentz
invariance of the exact theory do not necessarily spoil the Lorentz
invariance at low energy. The main difference with the Collins et
al. example is that they start with Lorentz violating propagators that do
not have the form that would result from a regularization of the
matter fields, something many expect LQG will provide.

We wish to thank Abhay Ashtekar, John Collins, Alejandro P\'erez,
Rafael Porto, Daniel Sudarsky, Luis Urrutia, Richard Woodard and
Nicol\'as Wschebor for comments. This work was supported in part by
grant NSF-PHY-0650715, funds of the Hearne Institute for Theoretical
Physics, CCT-LSU, Pedeciba and ANII PDT63/076.


\begin{thebibliography}{9}
\bibitem{loll}
  R.~Loll,
  Living Rev.\ Rel.\  {\bf 1}, 13 (1998)
  [arXiv:gr-qc/9805049]

\bibitem{spherical}
  R.~Gambini, J.~Pullin and S.~Rastgoo,
  Class.\ Quant.\ Grav.\  {\bf 26}, 215011 (2009)
  [arXiv:0906.1774 [gr-qc]].

\bibitem{laddha}
  A.~Laddha and M.~Varadarajan,
  Class.\ Quant.\ Grav.\  {\bf 27}, 175010 (2010)
  [arXiv:1001.3505 [gr-qc]].
\bibitem{oriti}
 D.~Oriti,
  ``Spin foam models of quantum spacetime'' Ph. D. Thesis 
[arXiv:gr-qc/0311066].

\bibitem{horava}
  P.~Horava, C.~M.~Melby-Thompson,
  Phys.\ Rev.\  {\bf D82}, 064027 (2010).
  [arXiv:1007.2410 [hep-th]].

\bibitem{collins}
  J.~Collins, A.~Perez, D.~Sudarsky, L.~Urrutia and H.~Vucetich,
  Phys.\ Rev.\ Lett.\  {\bf 93}, 191301 (2004)
  [arXiv:gr-qc/0403053].
\bibitem{collins2}
  J.~Collins, A.~Perez and D.~Sudarsky,
  in ``Approaches to quantum gravity'' D. Oriti, editor, Cambridge
  University Press (2009)
  [arXiv:hep-th/0603002].


\bibitem{smit}
J. Smit, ``Introduction to quantum fields on a lattice'', Cambridge
University Press (2002).


\bibitem{time}
  R.~Gambini, R.~A.~Porto, J.~Pullin and S.~Torterolo,
  Phys.\ Rev.\  D {\bf 79}, 041501 (2009)
  [arXiv:0809.4235 [gr-qc]].

\bibitem{rods}
R.~Gambini, R.~A.~Porto, J.~Pullin,
  Phys.\ Lett.\  {\bf A372}, 1213-1218 (2008).
  [arXiv:0708.2935 [quant-ph]].

\bibitem{doplicher}
 S.~Doplicher, K.~Fredenhagen, J.~E.~Roberts,
 Commun.\ Math.\ Phys.\ {\bf 172}, 187-220 (1995).  [hep-th/0303037].


\bibitem{reurve} 
C.~M.~Reyes, L.~F.~Urrutia, J.~D.~Vergara,
Phys.\ Rev.\ {\bf D78}, 125011 (2008); Phys.\ Lett.\ {\bf B675},
336-339 (2009).

\bibitem{abundant}
  W.~A.~Bardeen, M.~S.~Carena, S.~Pokorski and C.~E.~M.~Wagner,
  Phys.\ Lett.\  B {\bf 320}, 110 (1994)
  [arXiv:hep-ph/9309293];   A.~Brunstein, O.~J.~P.~Eboli and M.~C.~Gonzalez-Garcia,
  Phys.\ Lett.\ B {\bf 375}, 233 (1996) [arXiv:hep-ph/9602264].

\bibitem{pullin}
J. Pullin, talk presented at the GR19 meeting in Mexico City, July 2010.

\bibitem{Hossain:2010eb}
  G.~M.~Hossain, V.~Husain and S.~S.~Seahra,
  arXiv:1007.5500 [gr-qc].

\end{thebibliography}
\end{document}